\documentclass{ws-ijmpb}
\usepackage{amssymb,amsbsy,graphicx,float,subfigure}
\usepackage[latin1]{inputenc}

\begin{document}

\markboth{S. Piano et al.} {Point Contact Study of the
Superconducting Order Parameter in RuSr$_2$GdCu$_2$O$_8$}

%
\catchline{}{}{}{}{}
%

\title{Point Contact Study of the
Superconducting Order Parameter in RuSr$_2$GdCu$_2$O$_8$}

\author{S. PIANO$^*$, F. BOBBA, F. GIUBILEO, A. M. CUCOLO}
\address{Physics Department and INFM Supermat Laboratory, University of Salerno, \\
Via Salvatore Allende, I-84081 Baronissi (SA), Italy \\
$^*$samanta@sa.infn.it}

\author{A. VECCHIONE}
\address{Physics Department and INFM Coherentia Laboratory, University
of Salerno, \\
Via Salvatore Allende, I-84081 Baronissi (SA), Italy}
 \maketitle


\begin{abstract}
We have performed a detailed study of the conductance
characteristics obtained by point contact  junctions realized
between a normal Pt/Ir tip and syntered RuSr$_2$GdCu$_2$O$_8$
(Ru-1212) samples. Indeed, this compound is subject of great
interest due to the coexistence of both magnetic order and bulk
superconductivity. In our experiments, the low temperature
tunneling spectra reproducibly show a zero bias conductance peak
that can be well reproduced by a generalized BTK model in the case
of \emph{d-wave} symmetry of the superconducting order parameter.
\end{abstract}


\bigskip

From the discovery of high $T_c$ superconductors several
experimental and theoretical analysis have been realized to reveal
the electronic structure and to understand the pairing state of
these materials.
 Recently a new
compound of this family, the RuSr$_2$GdCu$_2$O$_8$ (Ru-1212), has
raised great interest. This rutheno-cuprate superconductor is a
triple perovskite material with alternating CuO$_2$ bilayers and
RuO$_2$ monolayers. In addition to the superconducting transition,
the rutheno-cuprate materials show magnetic order at $\sim 135 K$.
The magnetic order of the Ru moments is predominantly
antiferromagnetic along the $c$ axis\cite{Lynn}, while a
ferromagnetic component has been observed along the $a$, $b$
planes\cite{Bernhard2}. In this report we study the Point--Contact
conductance characteristics obtained on Ru-1212 compound. The
majority of the tunneling curves shows a zero-bias conductance
peak (ZBCP), which can be modeled by assuming a $d-wave$ pairing
symmetry of the superconducting order parameter in this material.

\begin{figure}[t!]
  \centering
  \includegraphics[width=6cm]{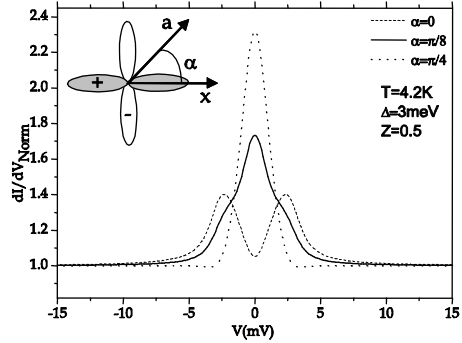}\\
  \caption{Numerical simulation of \emph{d-wave} BTK model at different values of angle $\alpha$ between
 the $x$-axis and $a$-axis of the superconductor (see inset). The fitting curve has been generated at $Z=0.5$,
 $T=4.2K$ and $\Delta=3meV$. The dashed line represents $\alpha=0$,
 the solid line $\alpha=\pi/8$ and the dotted line
 $\alpha=\pi/4$.}\label{teorico}
\end{figure}

 The Ru-1212 pellets were melt-textured synthesized samples\cite{Vecchione}.
 Resistivity measurements\cite{Attanasio} have shown a critical
 temperature $T_c^{onset}=40K$ and $T_c(\rho=0)=25K$. To realize our
 experiments we used a
Pt-Ir tip, chemically etched in a $37\%$ solution of HCl and in an
ultrasound bath. We have
 established the contact between the sample and the tip at $T=4.2K$, and we
measured the differential conductance $dI/dV$ as a function of
 the voltage bias $V$ using a conventional four-probe method and lock-in technique at $377Hz$.
We obtained  different conductance curves\cite{Cucolo}, majority
of which characterized by a triangular peak structure centered at
zero bias. Sometimes this peak is greater than 2, while in other
cases a variation of slope is present. These curves cannot be
explained by a \emph{s-wave} symmetry of the order
parameter\cite{Tesi}. For this reason we explore this feature in
term of a \emph{d-wave} symmetry. Indeed, \emph{d-wave} pairing
states
 have an internal phase of the pairing potential which
 modifies the surface states due to the interface effect of the
 quasiparticles (\emph{Andreev Bound
 States}). To take into account these effects, Tanaka and
 Kashiwaya\cite{Tanaka} modified the Blonder-Tinkham-Klapwijk (BTK)
 model\cite{BTK}, introducing a \emph{d-wave} symmetry of the order parameter.
 The BTK model describes the current-voltage characteristics  of a
  superconductor--normal metal junction separated by a
  barrier of arbitrary strength, which is modelled by a
dimensionless parameter $Z$: varying $Z$ it is possible
   to go by Andreev Reflection situation (small $Z$) to tunneling limit
   ($Z\gg1$). For an anisotropic \emph{d-wave} superconductor,
   at the given energy the tunnel current depends both on the
   incident angle $\varphi$ of electrons at the interface
   normal-superconductor as well as on the orientation $\alpha$, that is the angle
    of the order parameter
    $\Delta_\pm=\Delta\cos[2(\alpha\mp\varphi)]$.
   It is well known that in Point Contact experiments there is no preferential direction
    of the quasiparticle injection into the superconductor,
    so the tunneling current results by an integration over all directions inside a semisphere
   weigthed by the scattering probability term in the tunneling current formula. Moreover because our
   experiments deal with a policrystalline sample, more than one grain can be touched by
   the tip, consequently the angle $\alpha$ is a pure average fitting parameter,
   which depends on the experimental configuration.
\par
\begin{figure*}[t!]
\centering \subfigure[] {\includegraphics[width=5cm]{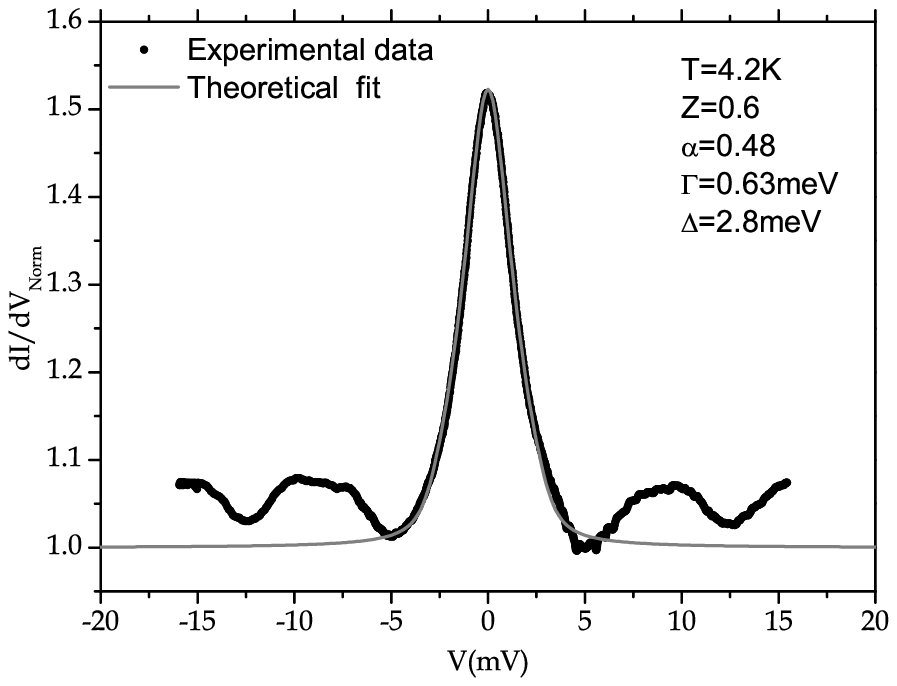}}
\subfigure[] {\includegraphics[width=5cm]{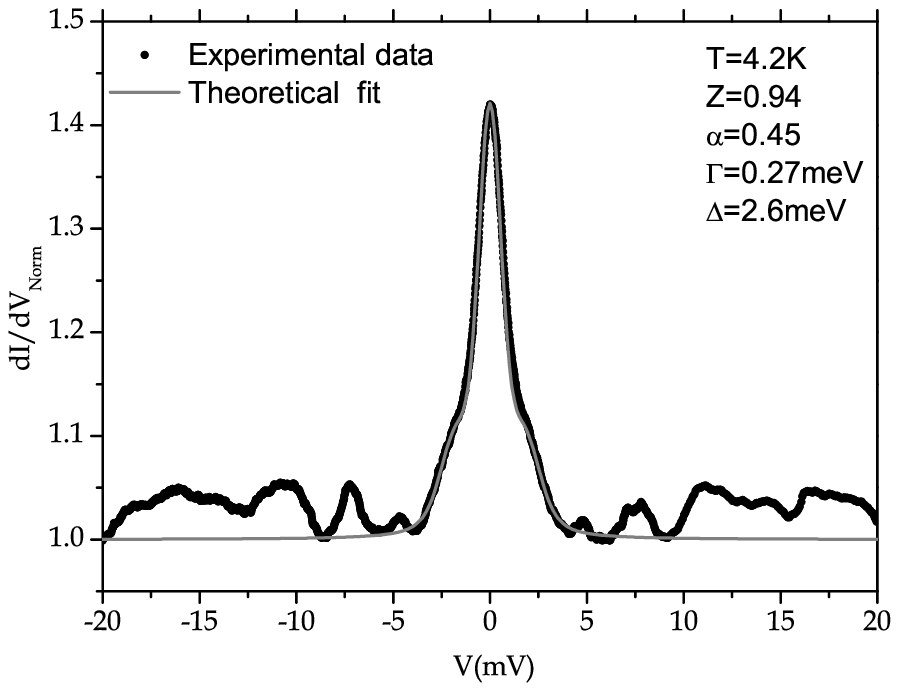}}
\caption{dI/dV-V characteristic measured on Ru-1212/Pt-Ir
point-contact junction at 4.2K. The points are experimental data.
The solid line is the fit of BTK model in $d-wave$, the
  fitting parameters are: (a) $\Delta=2.8meV$, $Z=0.6$, $\alpha=0.48$,
  $\Gamma=0.63meV$; $\Delta=2.6meV$, $Z=0.94$, $\alpha=0.45$, $\Gamma=0.27meV$. } \label{fit}
\end{figure*}

 In Fig.~\ref{teorico} we show an example of the effect of the
angle $\alpha$ on the shape of the
 conductance curves. We have fixed the barrier strength $Z=0.5$ and the temperature
 $T=4.2K$, and then we have assumed an order
 parameter, $\Delta$, of $3meV$.
 We note that, by varying $\alpha$, we obtain different curves. In particular
  for $\alpha=\pi/4$ the ZBCP
 reaches a maximum which is greater than 2, for $\alpha=\pi/8$ we
  note a shoulder in the ZBCP slope, and finally for $\alpha=0$
 we have not bound state formation being the ZBCP absent.
 To understand our experimental data and to investigate the symmetry
 of the order parameter in RuSr$_2$GdCu$_2$O$_8$, we have fitted the
 $dI/dV-V$ curves with the \emph{d-wave} modified BTK
 model\cite{Tanaka}.
 In Fig.~\ref{fit}, two experimental conductance curves are shown together
 with the best fitting curves.  Fitting parameters in our analysis are:
  the superconducting energy gap $\Delta$, the barrier strength $Z$, and
 the angle $\alpha$; in addition to these parameters, the smearing factor $\Gamma$ is introduced, which
 takes into account pair-breaking effects.
 The agreement between experimental data and theoretical model
is quite satisfactory, and it yields, for the magnitude of the order
parameter $\Delta$, a value ranging between $2.6meV$ and $2.8meV$.

In conclusion, our Point Contact spectroscopy experiments at
$T=4.2K$ on syntered RuSr$_2$GdCu$_2$O$_8$
 samples exhibit unconventional superconductivity.
 A peak is observed at zero bias in the $dI/dV-V$ curves due to
  the presence of Andreev Bound States at the surface.
 By fitting the experimental data with a \emph{d-wave} BTK
 model, we can estimate a
superconducting energy gap $\Delta=(2.7 \pm 0.1)\ meV$.


\smallskip



\begin{thebibliography}{99}

\bibitem{Lynn} J. W.  Lynn, B. Keimer, C. Ulrich, C. Bernhard and J. L. Tallon, Phys. Rev. B {\bf 61}, R14964 (2000).

\bibitem{Bernhard2} C. Bernhard {\it et al.},
Phys. Rev. B {\bf 59}, 14099 (1999).

\bibitem{Vecchione} M. Gombos, A. Vecchione, R. Ciancio, D. Sisti, S. Uthayakumar and S. Pace, Physica
C {\bf 408}, 189 (2004).

\bibitem{Attanasio} C. Attanasio, M. Salvato, R. Ciancio, M. Gombos, S. Pace, S. Uthayakumar and A. Vecchione,
Physica C {\bf 411}, 126 (2004).

\bibitem{Cucolo} F. Giubileo, F. Bobba, M. Gombos, S. Uthayakumar,
A. Vecchione, A. I. Akimenko, and A. M. Cucolo, Int. J. Mod. Phys. B
{\bf 17}, 3525 (2003).

\bibitem{Tesi} S. Piano, graduation thesis (University of Salerno, 2003).

\bibitem{Tanaka} S. Kashiwaya, Y. Tanaka, Rep. Prog. Phys. {\bf 63}, 1641
(2000).

\bibitem{BTK} G. E. Blonder, M. Tinkham and T. M. Klapwijk, Phys. Rev.
B {\bf 25}, 4515 (1982).

\end{thebibliography}
\end{document}